# Dynamics of higher-order solitons in regular and $\mathcal{PT}$-symmetric nonlinear couplers

R. Driben and B. A. Malomed

*Department of Physical Electronics, School of Electrical Engineering, Faculty of Engineering, Tel Aviv University, Tel Aviv 69978, Israel*

**Abstract.** Dynamics of symmetric and antisymmetric 2-solitons and 3-solitons is studied in the model of the nonlinear dual-core coupler and its PT-symmetric version. Regions of the convergence of the injected perturbed symmetric and antisymmetric $N$-solitons into symmetric and asymmetric quasi-solitons are found. In the PT-symmetric system, with the balanced gain and loss acting in the two cores, borders of the stability against the blowup are identified. Notably, in all the cases the stability regions are larger for antisymmetric 2-soliton inputs than for their symmetric counterparts, on the contrary to previously known results for fundamental solitons ($N$=1). Dynamical regimes (switching) are also studied for the 2-soliton injected into a single core of the coupler. In particular, a region of splitting of the input into a pair of symmetric solitons is found, which is explained as a manifestation of the resonance between the vibrations of the 2-soliton and oscillations of energy between the two cores in the coupler.

*Introduction.* The profound importance of optical solitons for fundamental studies and technological applications in photonics is well known [1]. In addition to the ubiquitous fundamental solitons, integrable models and physical media described by nearly-integrable equations [2] give rise to $N$-solitons, with $N \geq 2$, which are oscillating pulses periodically restoring their shape at distances that are multiples of the fundamental soliton period [3]. Experimentally, this was demonstrated for 2- and 3-order solitons in 1983 [4]. In a different experiment, initial narrowing of higher-order solitons was observed for values of $N$ up to 13 [5]. Higher-order solitons were also created in the cavity of a mode-locked dye laser operating at the wavelength of 620 nm [6]. Strongly oscillating higher-order solitons find natural applications for the pulse compression [7, 8] and frequency conversion [9]. A more recent, extremely important, application is the use of the fission of higher-order solitons as the source of ultra-broadband optical supercontinuum [10-13]. In particular, the enhanced nonlinearity of micro- and nano-structured materials may help to create higher-order solitons and catalyze their subsequent fission, using reduced pump intensities [14, 15].

In the present work we aim to study the dynamics of higher-order solitons in nonlinear dual-core couplers [16, 17] and their $\mathcal{PT}$ (parity-time)-invariant counterparts. The dynamics of fundamental solitons in couplers, the most important feature of which is the spontaneous symmetry breaking, i.e., a transition from symmetric solitons to asymmetric ones, has been analyzed in many works [18-21], the study of $N$-solitons being a natural extension of that analysis. *Inter alia*, one may expect a resonance between the frequency of the intrinsic vibrations of the higher-order solitons (which does not depend on $N$ [, ]) and the frequency of the field oscillations between two cores of the coupler. On the other hand, $\mathcal{PT}$-invariant counterparts of usual conservative systems are introduced as those with spatially separated balanced gain and loss, which admits the existence of a real spectrum and a continuous family of modes, instead of isolated states typical to generic dissipative systems [22]. It was proposed [23] and demonstrated experimentally [24] that $\mathcal{PT}$-balanced system can be readily built in optics, see also review [25].

Fundamental symmetric and antisymmetric solitons in the $\mathcal{PT}$-invariant version of the nonlinear coupler, with the balanced gain and loss applied to the two cores, were recently studied in Refs. [26-28]. Stability borders for such soliton families were found, in contrast to more general systems with imbalanced gain and loss acting in the separated cores, which give rise to isolated stable dissipative solitons in models of photonic and [29] and plasmonic [30] couplers (see a review in Ref. [31]), including two-dimensional solitons and vortices in planar couplers [32].

In this work, we identify stability borders for modes in the nonlinear coupler and its $\mathcal{PT}$-symmetric generalization, produced by symmetric and antisymmetric 2- and 3-soliton inputs, and compare the results with those found previously for fundamental solitons. Dynamical regimes for 2-solitons injected into one core of the coupler are investigated too. The analysis is performed by means of systematic simulations.

*Dynamics of higher-order solitons in nonlinear couplers*. The transmission of light in the lossless dual-core waveguide is described by the linearly coupled nonlinear Schrödinger (NLS) equations for amplitudes $u(z,t)$ and $v(z,t)$ in the two cores:

$$iu_z + (1/2)u_{tt} + |u|^2 u + v = 0,$$

$$iv_z + (1/2)v_{tt} + |v|^2 v + u = 0, \qquad (1)$$

where $z$ is the propagation distance and $t$ reduced time or transverse coordinate, in the temporal- or spatial-domain setting, respectively. Coefficients accounting for the dispersion or diffraction, Kerr nonlinearity, and inter-core coupling are all normalized here to be 1.

The temporal profiles of injected light corresponds to symmetric (+) or antisymmetric (−) $N$-solitons with fundamental amplitude $\eta$, $v(z=0,t) = \pm u(z=0,t) = N\eta \,\mathrm{sech}\,(\eta t)$.

In both cases, this input gives rise to obvious exact solutions to Eqs. (1), which are tantamount to the respective exact solutions of the single NLS equation. The issue is the stability of these solitons against perturbations that tend to destroy their symmetry or antisymmetry. This problem was tackled by means of systematic simulations of Eqs. (1), with the symmetry/antisymmetry-breaking perturbations introduced by adding 2% to and subtracting 2% from amplitudes of the two components. Such perturbations were found to be much stronger in affecting the solitons than other perturbation modes (for instance, symmetric ones, which act identically on both components). In principle, one may attack the stability problem differently, through the computation of Floquet multipliers for small perturbations around the time-periodic solutions [], but the implementation of such a rigorous analysis for the $N$-solitons is quite tricky.

Generic outcomes of the evolution of the perturbed symmetric and antisymmetric $N$-solitons for relatively low amplitudes are illustrated by Fig. 1. In this case, the higher-order solitons are unstable in the dual-core system, rearranging themselves into breather-like symmetric modes oscillating around fundamental solitons. Note that the breather generated by the antisymmetric input features a much larger amplitude of the intrinsic oscillations, but it always oscillates around a symmetric fundamental soliton, rather than an antisymmetric one. The latter finding illustrates the fact that, while antisymmetric fundamental solitons have a limited stability region [20], the symmetric solitons are more robust objects, which realize the ground state of the coupler [21]. With the increase of $\eta$, this scenario changes above a certain critical value $\eta_{cr}$, where the perturbed symmetric or antisymmetric higher-order soliton undergoes the spontaneous symmetry breaking, being transformed by the instability into an excited state oscillating around an *asymmetric* fundamental soliton, which is the basic propagation mode (ground state) in the nonlinear directional coupler for energies exceeding the respective threshold [18]-[21].

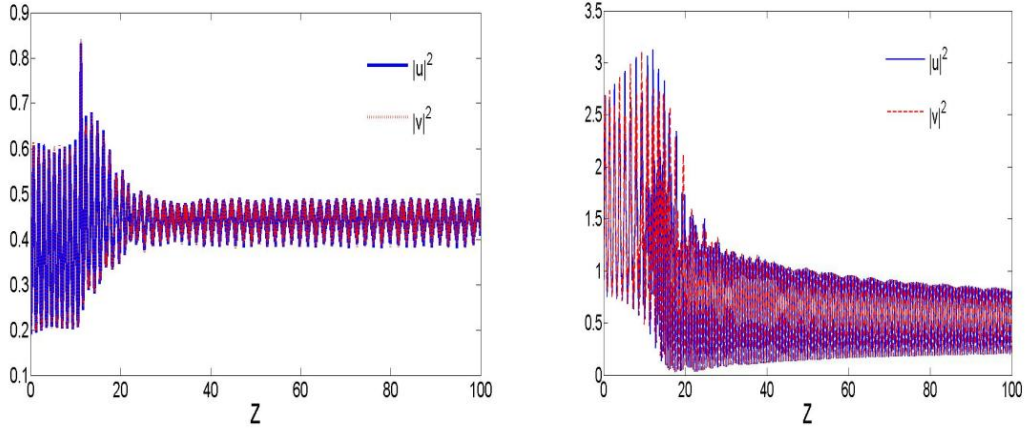

Figure 1: (Color online) (a) A typical example of the relaxation of a perturbed second-order symmetric soliton with $\eta^2=0.05$ into a persistent breather oscillating around a symmetric fundamental soliton. (b) An example of the convergence of a perturbed second-order antisymmetric soliton with $\eta^2=0.2$ into a symmetric breather.

These cases are shown, severally, in Figs. 2 and 3, for the symmetric 2-soliton with $\eta^2=0.2>\eta^2_{cr}\approx 0.090$, and for the antisymmetric one, with $\eta^2=0.5>\eta^2_{cr}\approx 0.366$. Note that, in both cases, the established mode features a strong asymmetry, with the pulse in one core being close to a fundamental soliton existing in this core without the coupling, while the weak component in the other core is a quasi-linear mode supported by its attraction to the strong component in the first core.

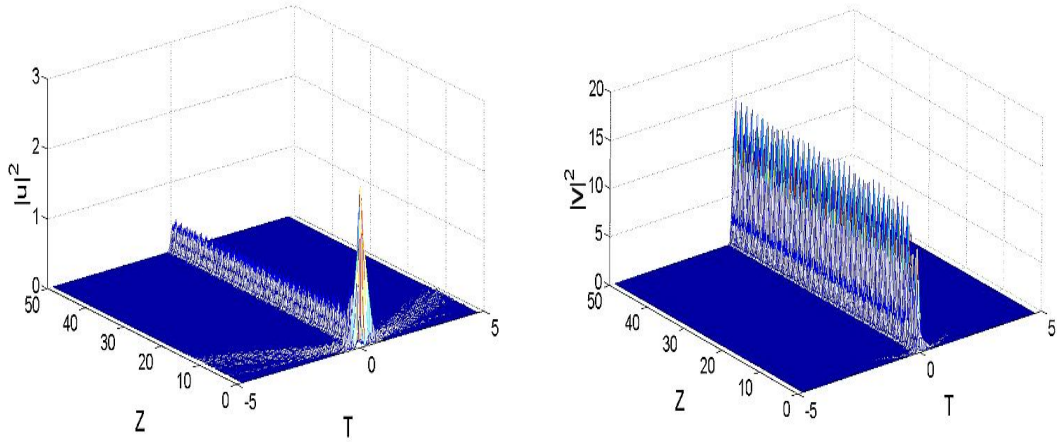

Figure 2: (Color online) A typical example of the relaxation of a perturbed second-order symmetric soliton with $\eta^2=0.2$ into a breather oscillating around a fundamental asymmetric soliton.

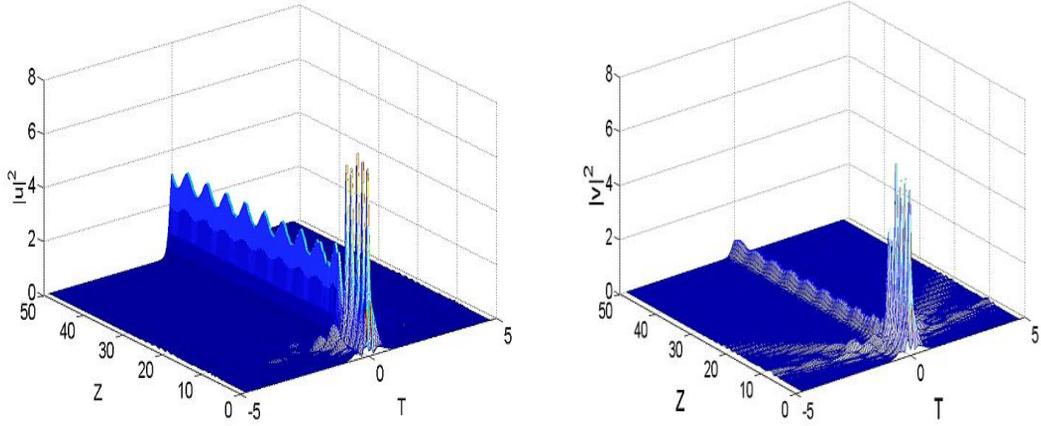

Figure 3: (Color online) An example of the convergence of a perturbed antisymmetric second-order soliton with $\eta^2$=0.5 into a strongly asymmetric breather.

| $N$ | 1 | 2 | 3 |
|---|---|---|---|
| symm | 4/3 | 0.090 | 0.018 |
| anti | 0.75 | 0.366 | 0.200 |

Table 1: The critical value of the squared amplitude $\eta^2$ for the symmetric and antisymmetric solitons of orders $N$ =1 (fundamental), 2, and 3, above which the soliton undergoes the spontaneous rearrangement into a breather oscillating around a fundamental asymmetric soliton.

The critical values for the spontaneous symmetry breaking are collected in Table 1. Included into the table are also $\eta^2_{cr}$=4/3 for the fundamental symmetric soliton, which is a well-known exact result [18], and $\eta^2_{cr} \approx 0.75$ for the antisymmetric soliton, which was found in a numerical form [20]. The critical values for the 2- and 3-solitons of both types, symmetric and antisymmetric, are new results obtained in the present work. It is worthy to note that $\left(\eta^2_{cr}\right)_{symm}$ rapidly decays with the increase of the soliton's order, and $\left(\eta^2_{cr}\right)_{anti}$ decays too, but much slower. As a result, while the threshold is higher for the fundamental ($N$=1) symmetric solitons than for their antisymmetric counterparts, the relation is *opposite* for $N$=2 and 3.

An essential characteristic of the dynamical rearrangement of the solitons of different types is the share of the initial energy, $\int_{-\infty}^{+\infty} \left[ |u(t)|^2 + |v(t)|^2 \right] dt$, which is kept by the established symmetric or asymmetric breather (despite the persistent vibrations, the breathers in the final state do not emit radiation). This share is shown in Fig. 4 as a function of $\eta^2$ for both the symmetric and antisymmetric 2-soliton inputs. A salient feature of the dependence is that the rearrangement of the antisymmetric 2-soliton into the symmetric or asymmetric mode gives rise to essentially larger loss than the rearrangement of its symmetric counterpart. This fact is natural, as the rearrangement is more dramatic for the antisymmetric input, always transforming it into a mode of the opposite symmetry. In fact, the strong loss suffered by the antisymmetric input explains its stronger effective stability in comparison with its symmetric counterpart, as the decrease of the pulse's energy pushes it farther from the instability border.

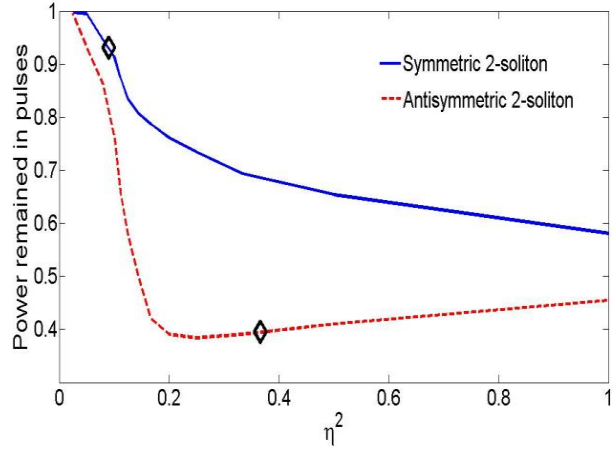

Figure 4: (Color online) The share of the total power kept in the final oscillating pulse as a function of input parameter $\eta^2$ for the initial symmetric and antisymmetric 2-solitons (the blue solid and red dashed curves, respectively). Marked are the threshold values $\eta^2_{cr}$ separating the transition into the symmetric and asymmetric breathers, see Table 1.

*Dynamics of higher-order solitons in PT-symmetric couplers*. The dual-core waveguide with the balanced gain and loss acting in the two cores is described by the following generalization of Eqs. (2) [26]:

$$iu_z + (1/2)u_{tt} + |u|^2 u - i\gamma u + v = 0, \qquad (2)$$

$$iv_z + (1/2)v_{tt} + |v|^2 v + i\gamma v + u = 0,$$

where $\gamma$ represents equal coefficients of the linear gain and loss. The PT symmetry holds in this system for $\gamma<1$. In this case, the inputs in the form of $v(t)=u(t)\exp\left(i\sin^{-1}(\gamma)\right)=N\eta\,\mathrm{sech}\,(\eta t)$, or $v(t)=-u(t)\exp\left(-i\sin^{-1}(\gamma)\right)=N\eta\,\mathrm{sech}\,(\eta t)$ automatically generate exact solutions to Eqs. (2) [26], in the form of $N$-solitons, which are, respectively, counterparts of the symmetric and antisymmetric solutions of Eq. (1). Accordingly, they may be called PT-symmetric and PT-antisymmetric modes.

For small $\eta^2$, both the PT-symmetric and antisymmetric inputs are transformed into breathers oscillating around PT-symmetric fundamental solitons. In this case, the loss and gain remain exactly balanced, hence the situation is not different from that considered above for the ordinary coupler, without the PT terms. However, unlike the regular coupler, the PT-symmetric system does not support asymmetric solitons, as the balance between the gain and loss is impossible for them, the symmetry/antisymmetry-breaking instability leading to blowup [26]. Thus, the basic issue is to identify the respective stability border, $\eta^2_{max}$. For the exact PT-symmetric and antisymmetric fundamental solitons, the borders were found in Ref. [26], in the analytical and numerical forms, respectively. Here, we have identified stability borders for the PT-symmetric and antisymmetric inputs in the form of 2-solitons, as shown in Fig. 5. For the sake of the comparison, the stability limits found in Ref. [26] for the fundamental solitons are shown too in the figure.

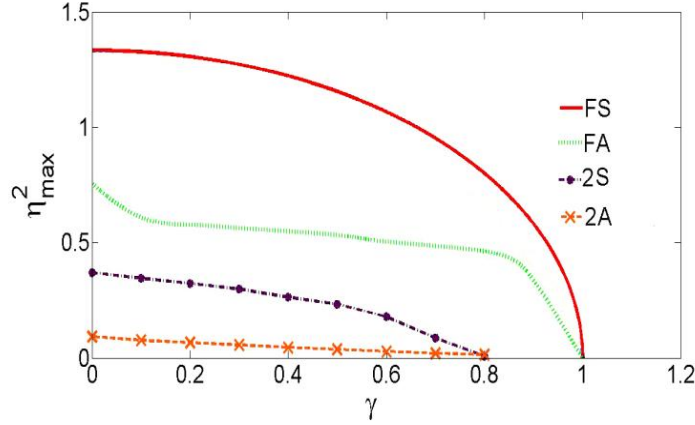

Figure 5: (Color online) Stability borders for the fundamental and second-order $\mathcal{PT}$-symmetric and antisymmetric solitons, in the $\mathcal{PT}$-balanced system. The input solitons with amplitude $\eta$ give rise to stable modes at $\eta^2 < \eta^2_{max}$. FS stands for the analytically found stability border for $\mathcal{PT}$-symmetric fundamental solitons, $\eta^2_{max} = (4/3)\sqrt{1-\gamma^2}$, and FA is the numerically found border for $\mathcal{PT}$-antisymmetric fundamental solitons [26]. New results are represented by the stability borders for the $\mathcal{PT}$-symmetric and antisymmetric 2-solitons (2S and 2A, respectively).

Note that the stability borders for the $\mathcal{PT}$-symmetric and antisymmetric 2-soliton inputs in Fig. 5 start, at $\gamma=0$, from values coinciding with those given for $N=2$ in Table 1. It is worthy to note that, similar to the situation in the system without the $\mathcal{PT}$ terms, the stability region is much smaller for the $\mathcal{PT}$-symmetric input than for its $\mathcal{PT}$-antisymmetric counterpart, on the contrary to the relation between the stability borders for the fundamental solitons.

*Switching dynamics of higher-order solitons in the nonlinear coupler.* Nonlinear couplers are most often used in the dynamical regime, launching the input signal (a fundamental soliton) into one core, and following oscillations of the energy between the cores. The most typical application of such regimes is nonlinearity-controlled switching between the cores [17]-[21], [34]. These studies suggest to investigate the dynamics of 2-solitons injected into a single core, i.e., simulate Eqs. (1) with initial conditions $u(z=0,t) = 2\eta\,sech(\eta t)$, $v(z=0,t)=0$, and compare results with those reported previously for the fundamental solitons.

For the input with large $\eta$ the nonlinearity dominates over the coupling, hence the 2-soliton stays, essentially, as the oscillating mode in the straight core, feeling little presence of the cross one. With the decrease of $\eta$, the amplitude of the oscillations decreases, maintaining a constant oscillation period. As $\eta$ approaches 1, this oscillating mode relaxes into a quasi-fundamental soliton coupled to a quasi-linear oscillating component in the cross core (similar to the picture displayed in Fig. 3). On the other hand, at small values of $\eta$, the input is converted into a breather with symmetric components in the two cores, oscillating around a fundamental soliton, which resembles the picture shown above for another situation (the antisymmetric 2-soliton input) in Fig. 1(b). A more interesting scenario occurs in the interval of $0.43 \leq \eta^2 \leq 0.67$, in the zone between the above-mentioned regimes of the strongly asymmetric and almost symmetric outputs. Close to the upper edge of this interval, around $\eta^2 = 0.67$, the frequency of the inner vibrations of the symmetric 2-soliton approaches the frequency of linear oscillations between the cores of the coupler.

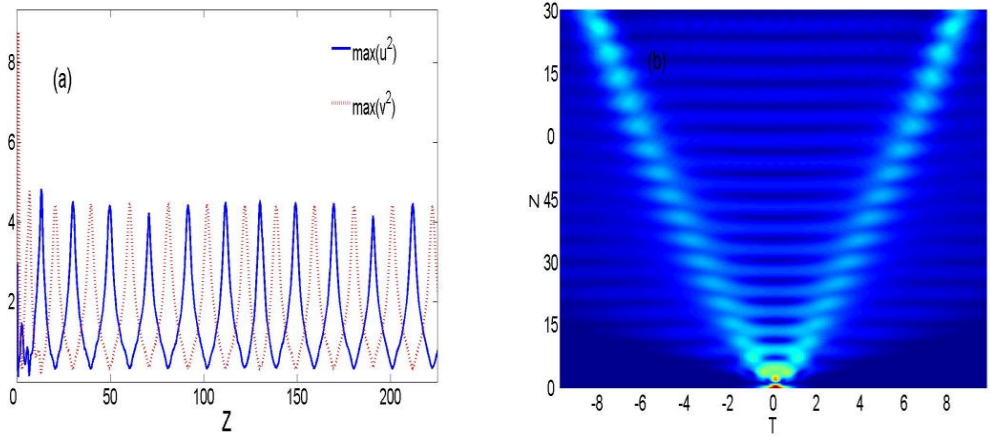

Figure 6: (Color online) (a) Periodic switching of light, coupled in the form of the 2-soliton into the single core, between the two cores of the coupler, at $\eta^2$=0.67. (b) Symmetric splitting of the 2-soliton, coupled into the single core, at $\eta^2$=0.53 (the field is shown in the straight core, the picture in the other one being almost identical).

As a result of the proximity to this resonance, a perfect periodic switching between the cores is observed, as shown in Fig. 6(a). A small decrease of $\eta^2$ gives rise to another effect, *viz.*, splitting of the initial 2-soliton into a pair of separating nearly symmetric fundamental solitons, as seen in Fig. 6(b).

*Conclusion.* We have studied the evolution of symmetric and antisymmetric higher-order solitons of orders *N*=2 and 3 in the basic model of the nonlinear coupler and in its *PT*-symmetric generalization. In the former model, we have identified regions of the convergence of injected perturbed *N*-solitons into symmetric and asymmetric breathers oscillating around fundamental solitons, symmetric or asymmetric ones. Critical values of input amplitude parameter, $\eta_{cr}$, which correspond to thresholds of the spontaneous symmetry and antisymmetry breaking, have been found for the 2- and 3-solitons. The evolution of the antisymmetric 2-soliton leads to a much larger radiation loss than in the case of the symmetric input. Asymmetric modes do not exist in the *PT*-symmetric system, with mutually balanced gain and loss acting in the linearly coupled cores. In this case, stability borders for the inputs in the form of *PT*-symmetric and antisymmetric 2-solitons have been found too. In all the cases, a noteworthy result is that the stability region for the antisymmetric 2-soliton is considerably larger than for its symmetric counterpart, on the contrary to the previously known results for fundamental solitons. This feature is explained by strong radiation loss of the rearranging antisymmetric mode, which pushes it back from the instability border. We have also studied the propagation of the 2-soliton injected into one core of the coupler (without the *PT* terms). In that case, different dynamical regimes are observed for different values of the input amplitude. In particular, a region of splitting of the input into a pair of symmetric solitons was found, which may be explained by proximity to the resonance between the intrinsic vibrations of the 2-soliton and field oscillations between the two cores.

# References

[1] Kivshar Y. S. and Agrawal G. P., *Optical Solitons: From Fibers to Photonic Crystals* (Academic Press, San Diego, 2003).
[2] Kivshar Yu. S. and Malomed B. A., Rev. Mod. Phys. **61** (1989) 763.
[3] Satsuma J. and Yajima N., Progr. Theor. Phys. (Suppl.) **55** (1974) 284.
[4] Stolen R. H., Mollenauer L. F., and Tomlinson W. J., Opt. Lett. **8** (1983) 186

[5] Mollenauer L. F., Stolen R. H., Gordon J. P., and W. J. Tomlinson, Opt. Lett. **8** (1983) 289.
[6] Salin F., Grangier P., Roger G., and Brun A., Phys. Rev. Lett. **56** (1986) 1132.
[7] Chan K. C. and Liu H. F., IEEE J. Quantum Electron., **31** (1995) 2226.
[8] Li Q., Kutz J. Nathan, and Wai P. K. A., **27** (2010) 2180.
[9] Lee K. S., Buck J. A., J. Opt. Soc. Am. B **20** (2003) 514
[10] Herrmann J., Griebner U., Zhavoronkov N., Husakou A., Nickel D., Knight J. C., Wadsworth W. J., Russell P. St. J., and Korn G., Phys. Rev. Lett. **88** (2002) 173901.
[11] Dudley J. M., Gentry G., and Coen S., Rev. Mod. Phys. **78** (2006) 1135.
[12] Skryabin D. V. and Gorbach A. V., Rev. Mod. Phys. **82** (2010) 1287.
[13] Driben R., Mitschke F., and Zhavoronkov N., Opt. Exp. **18** (2010) 25993.
[14] Driben R., Husakou A., and Herrmann J., Opt. Lett. **34** (2009) 2132.
[15] Driben R. and Zhavoronkov N., Opt. Exp. **18** (2010) 16733.
[16] Jensen S.M., IEEE J. Quantum Electron. **QE-18** (1982) 1580 ;
Maier A. M., Kvant. Elektron. (Moscow) **9** (1982) 3996.
[17] Romagnoli M., Trillo S. and Wabnitz S., Opt. Quant. Electron. **24** (1992) S1237.
[18] Wright E. M., Stegeman G. I., and Wabnitz S., Phys. Rev. A **40** (1989) 4455.
[19] Maimistov A. I., Sov. J. Quantum Electr., **21** (1991) 687.
[20] Soto-Crespo J. M. and Akhmediev N., Phys. Rev. E **48** (1993) 4710.
[21] Chu P. L., Malomed B. A., and Peng G. D., J. Opt. Soc. Am. B **10** (1993) 1379.
Malomed B. A., Skinner I., Chu P. L., and Peng G. D., Phys. Rev. E **53** (1996) 4084.
[22] Bender C. M., Rep. Prog. Phys. **70** (2007) 947.
[23] Ruschhaupt A., Delgado F., and Muga J. G., J. Phys. A **38** (2005) L171;
El-Ganainy R., Makris K. G., Christodoulides D. N., and Musslimani Z. H., Opt. Lett. **32** (2007) 2632.
Makris K. G., El-Ganainy R., Christodoulides D. N., and Musslimani Z. H., Phys. Rev. Lett. **100** (2008) 103904.
Klaiman S., Günther U., and Moiseyev N., Phys. Rev. Lett. **101** (2008) 080402.
Longhi S., Phys. Rev. Lett. **103** (2009) 123601.
[24] Guo A., Salamo G. J., Duchesne D., Morandotti R., Volatier-Ravat M., Aimez V., Siviloglou G. A., and Christodoulides D. N., Phys. Rev. Lett. **103** (2009) 093902;
Ruter C. E., Makris K. G., El-Ganainy R., Christodoulides D. N., Segev M., and Kip D., Nature Phys. **6** (2010) 192.
[25] Makris K. G., El-Ganainy R., Christodoulides D. N., and Musslimani Z. H., Int. J. Theor. Phys. **50** (2011) 1019.
[26] Driben R. and Malomed B. A., Opt. Lett. **36** (2011) 4323.
[27] Driben R. and Malomed B. A., EPL **96** (2011) 51001.
[28] Alexeeva N. V., Barashenkov I. V., Sukhorukov A. A., and Kivshar Y. S., Phys. Rev. A **85** (2012) 063837.
[29] Malomed B. A. and Winful H. G., Phys. Rev. E **53** (1996) 5365;
Atai J. and Malomed B. A., Phys. Rev. E **54** (1996) 4371; Phys. Lett. A **246** (1998) 412;
Sakaguchi H. and Malomed B. A., Physica D **147** (2000) 273;
Firth W. J. and Paulau P. V., Eur. Phys. J. D **59** (2010) 13.
[30] Marini A., Skryabin D. V., and Malomed B. A., Opt. Exp. **19** (2011) 6616.
[31] Malomed B. A., Chaos **17** (2007) 037117.
[32] Paulau P. V., Gomila D., Colet P., Loiko N. A., Rosanov N. N., Ackemann T., and Firth W. J., Opt. Exp. **18** (2010) 8859;
Paulau P. V., Gomila D., Colet P., Malomed B. A., and Firth W. J., Phys. Rev. E **84** (2011) 036213.
[33] Murdock J. A., *Perturbations: Theory and Methods* (SIAM: Philadelphia, 1999).
[34] Chu P. L., Malomed B. A., Peng G. D., and Skinner I., Phys. Rev. E **49** (1994) 5763;
Skinner I. M., Peng G. D., Malomed B. A., and Chu P. L., Opt. Commun. **113** (1995) 493.